\documentclass{epl}
\usepackage{epsfig,amssymb,amsfonts,amsmath}
\title{Dynamics of dilute disordered models: a solvable case}
\shorttitle{Dynamics of dilute disordered models}
\author{Guilhem Semerjian\inst{1}\thanks{E-mail: \email{guilhem@lpt.ens.fr}} \and Leticia F. Cugliandolo\inst{1,2}\thanks{E-mail: \email{leticia@lpt.ens.fr}}}
\shortauthor{G. Semerjian and L.F. Cugliandolo}
\institute{
  \inst{1} Laboratoire de Physique Th{\'e}orique de l'Ecole Normale
Sup{\'e}rieure - 24 rue Lhomond, 75231 Paris Cedex 05, France\\
  \inst{2} Laboratoire de Physique Th{\'e}orique et Hautes Energies -
4 place Jussieu, 75252 Paris Cedex 05, France
}
\pacs{75.10.Nr}{Spin-glass and other random models}
\pacs{75.10.Hk}{Classical spin models}
\pacs{64.70.Pf}{Glass transitions}

\begin{document}

\maketitle

\begin{abstract}
We study the dynamics of a dilute spherical model with 
two body interactions and random exchanges. We analyze the 
Langevin equations and we introduce a functional 
variational method to study generic 
dilute disordered models. A crossover temperature replaces 
the dynamic transition of the fully-connected limit. There are two asymptotic 
regimes, one determined by the central band of the spectral density 
of the interactions and a slower one determined 
by localized configurations on sites 
with high connectivity. We confront the behavior of this model
to the one of real glasses.
\end{abstract}

Kirkpatrick, Thirumalai and Wolynes showed that 
a family of fully connected ({\sc fc}) disordered spin systems realize 
schematically the phenomenology of the 
glass transition~\cite{KTW}. These models are
very useful for several reasons: (i) their statics and dynamics are 
solvable with a variety of techniques; (ii) since their macroscopic
dynamic equations are identical to those stemming from self-consistent 
approximations to more realistic interacting systems (e.g.
mode-coupling ({\sc mca})~\cite{Gotze})
they limit the range of validity of the latter; (iii) 
their defects are clear ({\sc fc}, infinite 
dimensions) and some improvements needed in a
better description of real systems can be identified. 
The {\sc fc} p spin 
disordered models are members of this 
group.
Models of finite dimensional manifolds embedded in infinite dimensions
under quenched random potentials are generalizations 
that also yield real space information~\cite{Bou-review}.

The main features of the {\sc fc} 
spherical p spin model, on which we
focus for concreteness, are the following. 
(i) It has a
dynamic transition at $T_d<+\infty$. Above $T_d$ an 
infinite system ($N\to\infty$) equilibrates with its environment~\cite{KTW}. 
Below $T_d$ the equilibration time diverges with $N$~\cite{Cuku}. 
Right 
above $T_d$ the dynamics
coincides~\cite{KTW} with the schematic {\sc mca} to super cooled
liquids~\cite{Gotze} and correlations decay in two steps. 
Below $T_d$ for any waiting-time $t_w$ two-time functions
decay to zero after a sufficiently long
subsequent time, with aging effects~\cite{Cuku}.
(ii) It has a rich structure of metastable ({\sc tap})  
states that are stationary points of a
free-energy density $f$ function of  the
relevant order parameters ($m_i \equiv \langle s_i\rangle$
in this case)~\cite{Kupavi,Cagipa}. 
Within 
$(f_{\sc min}, f_{\sc th}]$ the number of metastable states, ${\cal N}(f)$, 
is exponential in $N$ leading to a 
finite complexity $\Sigma(f) \equiv N^{-1} \ln {\cal N}(f)$. 
($f_{\sc min} < f_{\sc th}$ when $p\geq 3$.)
Strictly below $f_{\sc th}$ minima are exponentially dominant
with respect to saddles.
Typically, the states 
on the threshold ({\sc th}) are marginal~\cite{Kupavi}
and attract the dynamics when times diverge after $N\to\infty$~\cite{Cuku,Bi}.
(iii) At $T_d$ the liquid state
($m_i=0 \, \forall i$) is fractured into an exponential in $N$ number 
of {\sc tap} states with $m_i\neq 0$. 
At $T_s$ the partition sum is dominated by the
lowest states with $f_{\sc min}=f_{\sc eq}$ and, since 
$\Sigma(f_{\sc min})=0$, there is an
entropy crisis associated to the Kauzmann paradox~\cite{KTW}.
(iv) The barriers between {\sc tap} 
states are expected to be 
$O(N)$ leading to a separation in the time-scales 
to descend below the threshold.

In finite dimensional systems with finite range interactions,
however, the dynamic transition is only a crossover
at $T_g$ and it is not clear if the 
static transition exists at all. 
The dynamics must penetrate below the threshold to
slowly descend towards equilibrium.
Within this picture two distinct regimes would appear in the
low-$T$ isothermal dynamics of real systems: a {\sc fc}-like one when the 
system approaches a pseudo-threshold of flat directions in phase space
and a slower activated regime  in which  the system jumps 
over barriers to relax its excess energy density and very slowly progress
towards equilibrium~\cite{Bou-review}. 
This view explains the cooling-rate effects 
since crossing $T_g$ with a slower rate allows to penetrate deeper
below the threshold. 
How and if the aging properties in the first and second regime
resemble is a very interesting open problem. One could
attack it analytically by studying the dynamics of p spin 
models in time scales that diverge with $N$ though this is very hard. 
Simulations of finite size {\sc fc}  models~\cite{Crri} 
and Lennard-Jones mixtures~\cite{LJ,LJ2} support the existence of 
the two aging regimes.

In this Letter
we study an alternative model that realizes explicitly two
nonequilibrium dynamic regimes,
a {\sc fc}-like one and a slower one that we relate to transitions between 
metastable states with {\it finite} life-time.
The merits of the model are 
manifold. Its dynamics is tractable analytically.
Being defined on a random graph the time-scale dependent 
metastable states are determined by the ``real space geometry'':
they are localized configurations on sites with high connectivity. 
The model has also some drawbacks that we discuss after 
presenting its solution. Still, 
we expect to use it as a starting point for the study of more
realistic cases.

We solve the dynamics of the dilute spherical model:
\begin{equation}
H_J[\vec s] = - \frac{1}{2} 
\sum_{i \neq j} J_{ij} s_i s_j 
\; .
\label{hamiltonian}
\end{equation}
where the spins are continuous unbounded variables, spherically constrained
$\sum_{i=1}^N s_i^2=N$.
The couplings $J_{ij}$ are quenched random variables, 
independently and identically distributed with
\begin{eqnarray}
P(J_{ij}) &=& \left(1- p/N \right) \, \delta(J_{ij}) + p/N \, \pi (J_{ij})
\: , \nonumber \\
\pi (J_{ij}) &=&\frac12  \; 
[\delta(J_{ij}-1/\sqrt{p}) + \delta(J_{ij}+1/\sqrt{p})]
\; ,
\end{eqnarray}
 that satisfy $J_{ii}=0$ and $J_{ij}=J_{ji}$.
When $p$ is finite and $N$ goes to infinity, this distribution law
generates a random graph, so
that the connectivity of a given site is distributed according to a Poisson law
with average $p$. Contrary to the {\sc fc} case, i.e. the complete
graph obtained in the $p =N \to \infty$ limit, each spin interacts with a
finite number of ``neighbors'', as in a finite dimensional system. Yet,
the ``neighbors'' are chosen randomly among the other spins and 
there is no geometrical distance. The model is of mean-field type. 
In all the following we take $p \gg  1$ but finite with respects to $N$.
The spins evolve with the Langevin equation 
\begin{equation}
\dot s_i(t) = - \delta H_J[\vec s]/\delta s_i(t) - z(t) s_i(t) +\xi_i(t)
\label{langevin}
\end{equation}
where the Lagrange multiplier $z(t)$ enforces the 
spherical constraint and $\xi_i(t)$ is a white noise with zero mean and 
variance $\langle \xi_i(t) \xi_j(t_w) \rangle = 2T \delta_{ij} \delta(t-t_w)$.
($k_B=1$ and we rescaled time to absorb the friction coefficient.)

We solve the model in two ways. By rotating the spin
$\vec s=(s_1,\dots,s_N)$ onto the basis of eigenvectors of the 
random matrix $J_{ij}$ the Langevin equations of any
continuous quadratic model decouple and can be completely 
solved~\cite{spherical-dyn,Cude}. 
Many interesting dynamic observables depend only on 
the density of states $\rho(\mu  )$ of the interaction matrix $J_{ij}$. We 
estimate their scaling in different regimes analytically and 
we confront them to the numerical solution of the 
dynamical equations where we approximate $\rho(\mu  )$ 
with Eq.~(29) in \cite{Secu}.
Then we introduce 
a more general method that can also be applied to non-quadratic
models and/or soft spins.
It is a dynamic generalization of the iterative method
used to find $\rho(\mu  )$ for 
dilute random matrices~\cite{Bimo}. 

For future reference, 
let us summarize the main properties of 
$\rho(\mu  )$~\cite{density-of-states,Secu}. 
It has a symmetric central band in
$[-\lambda_c(p),\lambda_c(p)]$, a crossover 
extending beyond $|\lambda_c(p)|$
that is not known in detail, and two tails that 
vanish as $\rho(\mu  ) \sim \exp[-p \mu  ^2\ln \mu  ^2]$ 
when $\mu  \to\pm  \infty$. The tails 
are due to large fluctuations of
the local connectivity. For $k\gg  2p$ a site with $k$ neighbors
gives rise to an eigenvalue $\mu  \sim \sqrt{k/p}$ with a localized 
eigenvector $\vec v_\mu  $ on it. When $p\to\infty$, $\lambda_c(p)\to 2$
and the tails disappear.

At time $t$ after preparation in the initial condition $s_\mu  (0)$ 
the rotated spin $s_\mu  (t)= \vec v_\mu  \cdot  \vec s(t)$ 
is given by~\cite{Cude}
\begin{eqnarray*}
s_\mu  (t) \sqrt{\Gamma(t)}= s_\mu  (0) e^{\mu  t} +\int_0^t dt' 
\, e^{\mu  (t-t')} \sqrt{\Gamma(t')} \xi_\mu  (t') 
\; .
\end{eqnarray*}
$\Gamma(t) \equiv \exp[2\int_0^t dt' z(t')]$ is 
related to the energy density, 
$\epsilon(t) = T/2- d_t \ln \Gamma(t)/4$,
and it determines the decay of the 
self-correlation $NC(t,t_w) = 
\sum_{i=1}^N [\langle s_i(t) s_i(t_w) \rangle]_J$
and linear self-response $NR(t,t_w) = 
\sum_{i=1}^N \delta [\langle s_i(t) \rangle]_J /\delta h_i(t_w)|_{h=0}$.
We choose the initial condition $s_\mu  (0)=\pm  1 \, \forall \mu  $ that
mimics an instantaneous quench from 
$T=+\infty$ at  $t=0$.
Imposing the spherical constraint $C(t,t)=
\int d\mu  \rho(\mu  ) \langle s^2_\mu  (t) \rangle =1$ implies
\begin{equation}
\Gamma(t) = f(t) + 2T \int_0^t dt' \; f(t-t') \Gamma(t')
\; ,
\label{volterra}
\end{equation}
with 
$f(t) \equiv \int d\mu \,  \rho(\mu  ) e^{2 \mu  t}$. 
Asymptotically $f(t)$ can be estimated 
with a saddle-point evaluation of the integral.
When it is dominated by $\mu  \sim \lambda_c(p)$, 
as in the {\sc fc} limit,
$f(t) \sim \exp[2\lambda_c(p) t]/t^{3/2}$. After a 
crossover time $t^{0}_{\sc co}(p)$
the integral  is dominated by $\mu  $'s in the crossover 
in $\rho(\mu  )$  and $f(t)$ grows faster than an exponential 
though we cannot determine
its functional form. Finally, when times are so long as to explore 
the tails of $\rho(\mu  )$, $f(t) \sim \exp[t^2/(2 p \ln t)]$. 

\begin{figure}
\centerline{
\epsfig{file=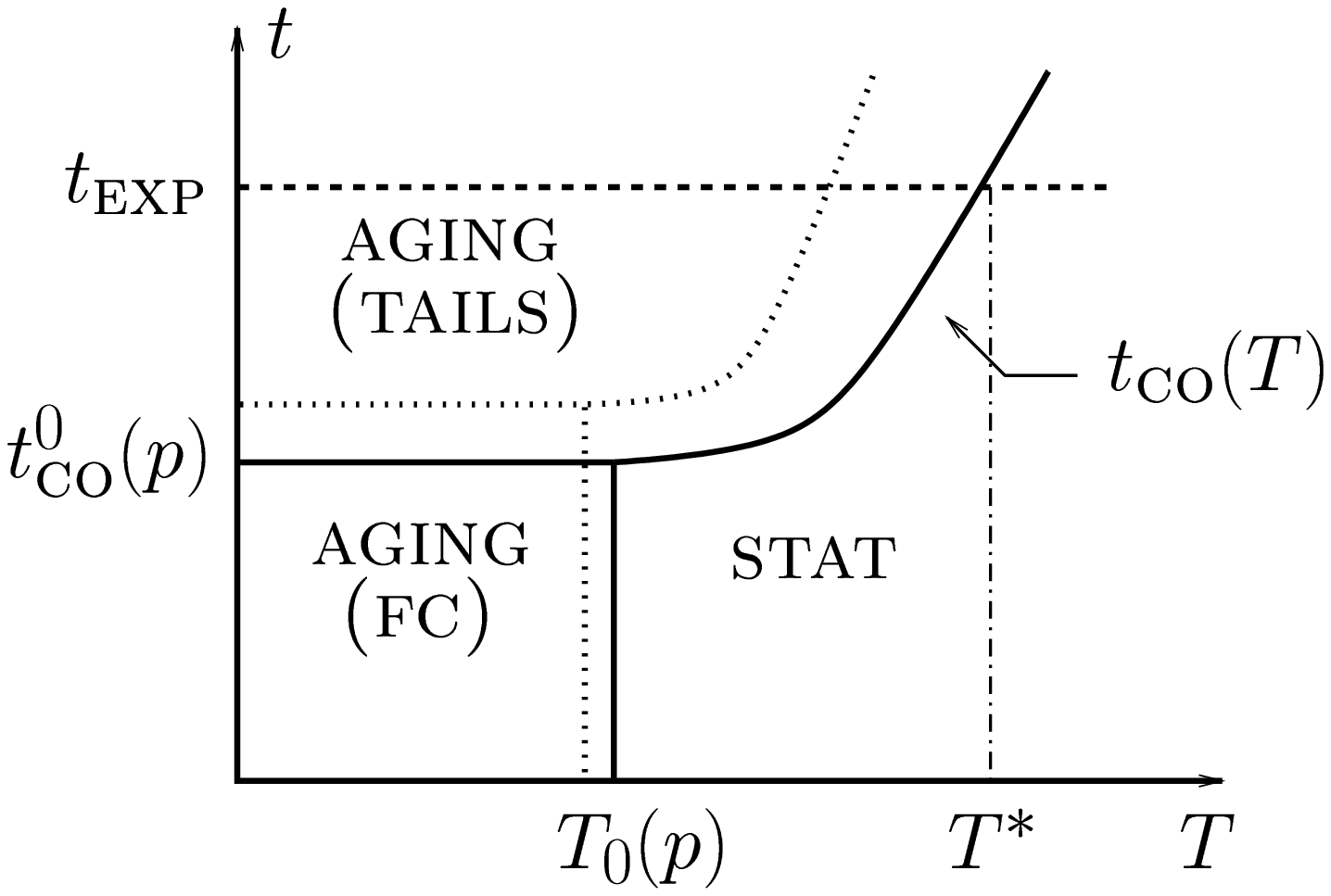,height=4.2cm}
\hspace{0.25cm}
\epsfig{file=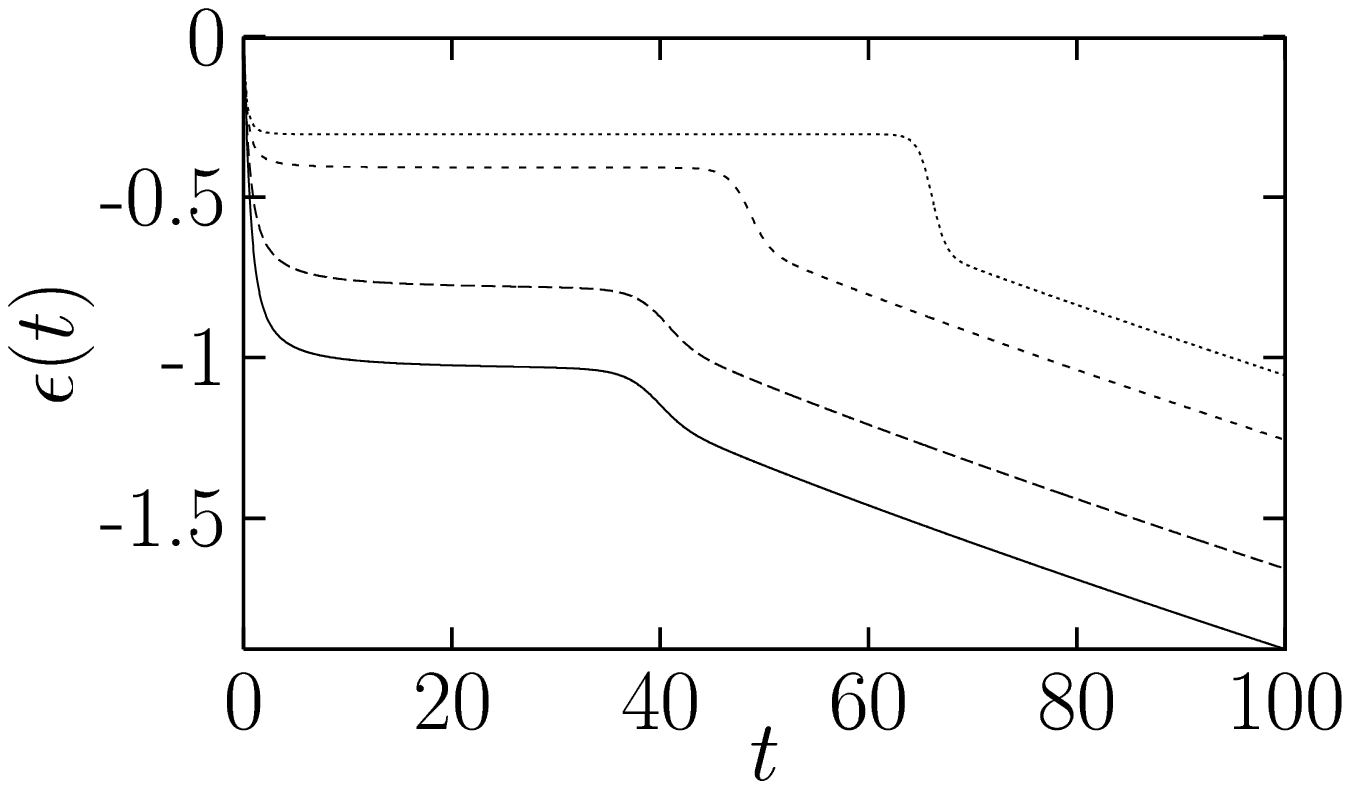,height=4.2cm}
}
\vspace{0.25cm}
\caption{Left: the crossover time as a function of temperature. Solid 
line $p=p_1$, dashed line $p=p_2$. $p_1<p_2$. Right:
Decay of the energy-density. $p=10$. From bottom to top:
$T=0$, $T=0.5$, $T=1.3$ and $T=1.7$.
}
\label{sketch}
\end{figure}

We thus have a first crossover time $t^{0}_{\sc co}(p)$ where the behavior of
$f$ changes from exponential (with polynomial corrections) to faster
than exponential. The more interesting function $\Gamma(t)$ is
determined from $f(t)$ by Eq.~(\ref{volterra}). Note that at $T=0$,
these two functions coincide. When $T$ is raised to a positive value,
$\Gamma(t)$ has still a
crossover time $t_{\sc co}(T,p)$ beyond which it grows faster than
exponential. The important feature which comes out of
Eq.~(\ref{volterra}) is that $t_{\sc co}(T,p)$ remains close to
$t^{0}_{\sc co}(p)$, its zero temperature value, until a crossover
temperature $T_0(p)$ is reached. For higher temperatures $t_{\sc
co}(T,p)$ grows with $T$. These time-temperature regimes are sketched 
in the left panel of 
Fig.~\ref{sketch}. More quantitatively, this behavior amounts to:

\noindent
--- when $T<T_0(p)$:
$\Gamma(t) \sim f(t)/(1-T/T_0(p))^2$ for $1\ll  t <  t_{\sc co}^0(p)$
and $\Gamma(t)\sim f(t)$ for $t\gg  t_{\sc co}^0(p)$. 

\noindent
--- when $T> T_0(p)$:  
$\Gamma(t) \sim \exp[b(T) t]$ for $1\ll  t <  t_{\sc co}(T,p)$
and
$\Gamma(t)\sim f(t)$ for $t\gg  t_{\sc co}(T,p)$
with $t_{\sc co}(T,p)$ a growing function of $T$.

$T_0(p)$ is the trace
of the phase transition in the {\sc fc} limit ($T_0(\infty)=1$).
It separates a low $T$ regime in which the 
model has a {\sc fc}-like out of equilibrium 
decay for long time-scales and a much 
slower one for still later epochs, from a high $T$ regime where the 
system crosses over from an equilibrium-like behavior 
at long time-scales to the tail dominated 
very slow out of equilibrium decay at much longer time-scales. 
For all $T$ the system eventually reaches the tail-dominated regime
but the higher the temperature $(>T_0(p))$ the longer the equilibrium-like 
regime lasts.

The time-scales just described can also be visualized from the 
decay of $\epsilon(t)$ displayed in the right panel in 
Fig.~\ref{sketch}. 
For the two lower curves 
$T<T_0(p)\sim 1.1$ and $t_{\sc co}^0(p) \sim 35$. The 
plateau occurs at $[T-\lambda_c(p)]/2$ as in the {\sc fc}
limit. For the two upper curves $T>T_0(p)$ and $t_{\sc co}(T,p)$
increases with $T$. For all $T$, when $t \gg  t_{\sc co}(T,p)$, we have  
$\epsilon(t) \sim -t/(4 p \ln t)$ meaning that the energy density 
diverges asymptotically, a result that is consistent with an
unbounded $\rho(\mu  )$. The 
dynamics in the tail-dominated regime is independent of $T$, so not
activated.
 
We explain in this paragraph how these results can be viewed as a
sketchy picture of a glass ``transition'', see the left panel of
Fig.~\ref{sketch}. 
Imagine that experimentally one 
has access to times that are shorter than $t_{\sc exp}$, choose a
value of $p$ such that $t_{\sc co}^0(p) < t_{\sc exp}$ and define
$T^*$ by $t_{\sc exp}=t_{\sc co}(T^*,p)$. $T^*$ thus depends on the
experimental time-scale available, and is larger than $T_0(p)$.
When $T<T_0(p)$ and times are
shorter than $t_{\sc co}^0(p)$ the system ages as in the {\sc fc} limit
while beyond $t_{\sc co}^0(p)$ aging is of a different
kind, dominated by the tails of the distribution. 
When $T>T^*$ all accessible times 
are shorter than $t_{\sc co}(T,p)$ and the dynamics is stationary as in the 
high-$T$ phase of the 
{\sc fc} limit. (Note however that these results are not 
accessible with a 
naive static calculation since $\epsilon$ is not bounded from 
below.) Finally, there is an intermediate temperature regime,
$T_0(p) < T < T^*$ in which the dynamics crosses 
over from stationary to tail dominated. The width of this ``unnatural'' regime 
can be shrinked by using larger $p$'s since 
$T_0(p)$ very weakly decreases with $p$ while 
$t_{\sc co}^0(p)$ grows with $p$, cfr.~the two curves
in the left panel of Fig.~\ref{sketch}. One concludes that the dynamics 
behaves similarly to what expected in real glasses in the sense that 
within the experimentally limited time-window 
there are two aging regimes when $T< T_0(p)$ and a single stationary
regime at high $T$. 

From now on, in order to shorten the notation
we call $T_0$ the crossover temperature and 
$t_{\sc co}$ the crossover time at the considered
$T$ and $p$, i.e. $t_{\sc co}=t_{\sc co}^0(p)$ if $T<T_0(p)$
and 
$t_{\sc co}=t_{\sc co}(T,p)$ if $T>T_0(p)$.
When $p\to\infty$, 
$t_{\sc co}$ moves to infinity for all $T$, $\rho(\mu  )$
approaches the semi-circle law and $\Gamma(t)$ takes the {\sc fc} form.

The evolution in the {\sc fc} limit was easy to grasp by following 
$\langle s_\mu  (t)\rangle$ where the low-$T$ non-equilibrium 
dynamics corresponds to an 
incomplete condensation on $\mu  =2$~\cite{Cude}. 
Let us focus on $T<T_0$ and analyze 
$\langle s_\mu  (t)\rangle$ in the dilute case.
When $t$ is in the {\sc fc}-like regime, the projections on
all eigenvectors with $\mu  <\lambda_c(p)$ are exponentially suppressed,
the one with $\mu  =\lambda_c(p)$ increases as a power law in time and the ones
with $\mu  >\lambda_c(p)$ increase exponentially in time. 
When $t$ enters the 
tail-dominated regime, the cut-off
$\mu  _{\sc c}(t)$ which separates the eigenvectors that are 
exponentially suppressed [$\mu  <\mu  _{\sc c}(t)$] 
from those that increase exponentially [$\mu  >\mu  _{\sc c}(t)$] moves
away from $\lambda_c(p)$ and eventually grows as $\mu  _{\sc c}(t) \sim
t/(4 p \ln t)$. In real space
this means that localized states on all sites with connectivities
($k> p \mu  _{\sc c}^2$) increase exponentially for 
times $<t(\mu  _{\sc c})$ while all others are exponentially 
suppressed. At longer times sites with higher connectivities dominate.

\begin{figure}
\centerline{
\epsfig{file=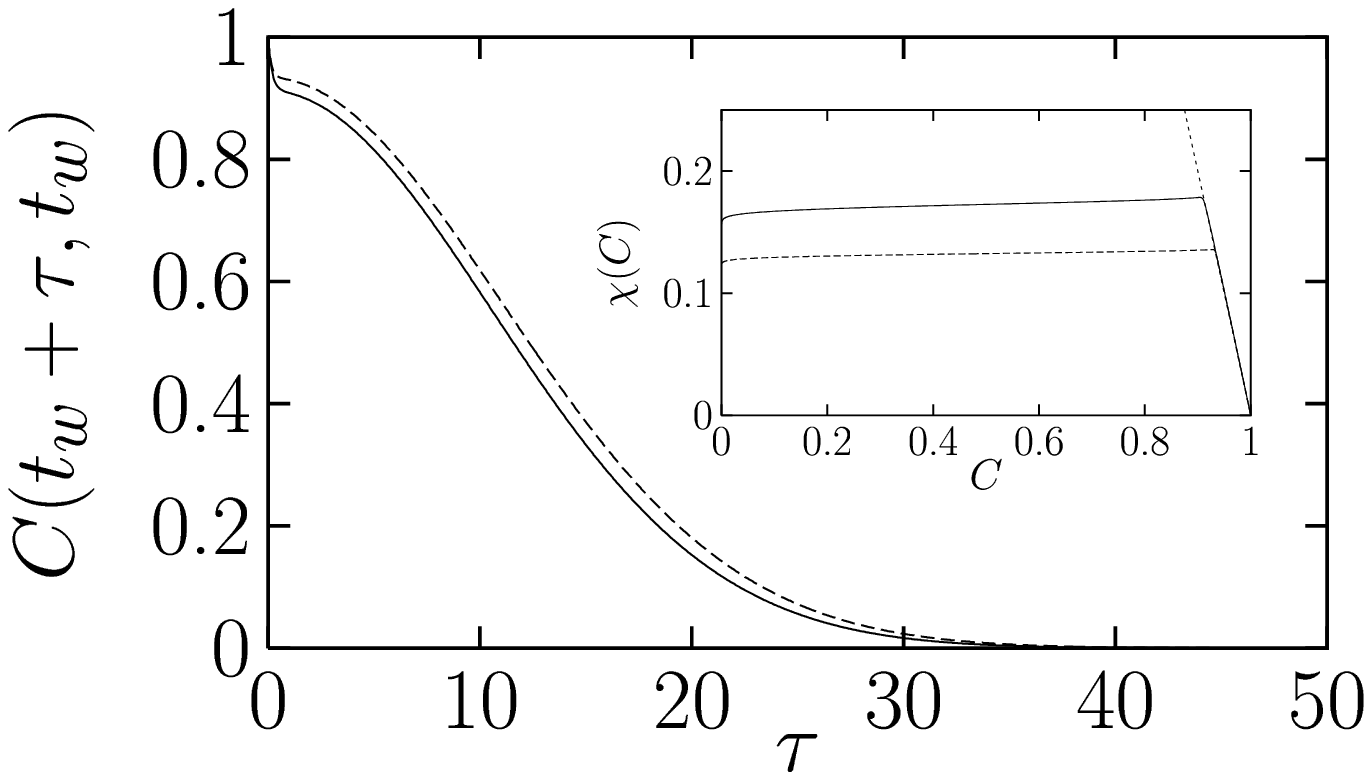,height=3.8cm}
\epsfig{file=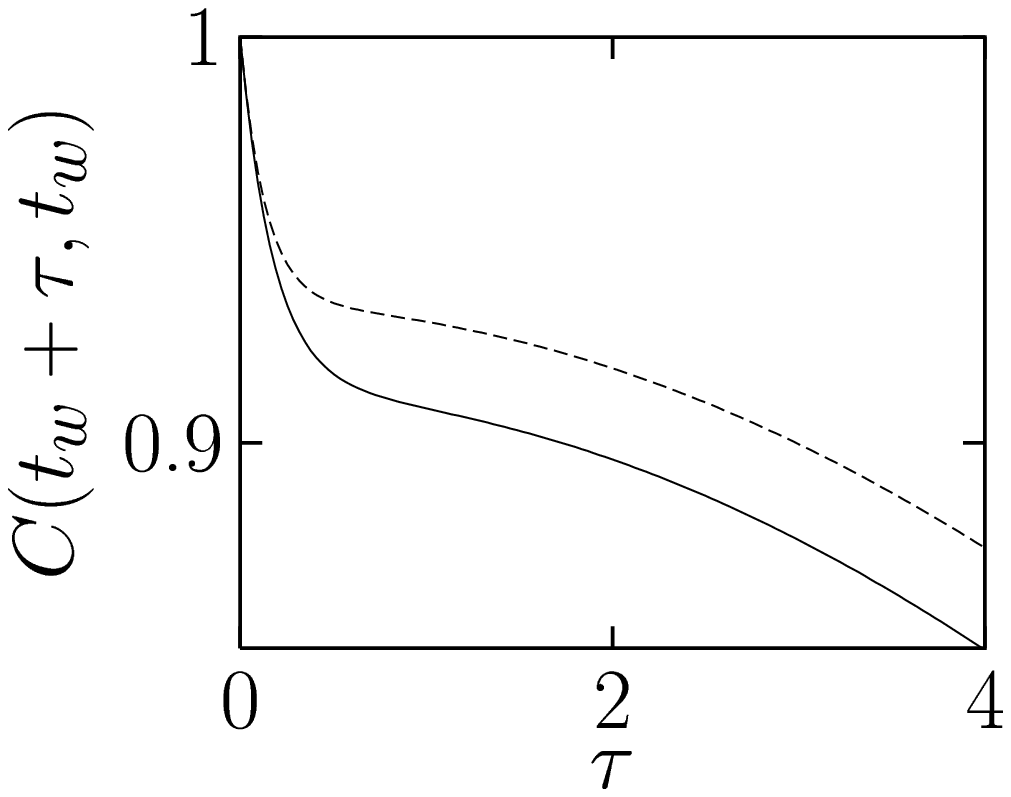,height=3.8cm}
}
\vspace{0.2cm}
\caption{Left: The self-correlator in the tail dominated regime
for $t_w=200$ (solid) and $300$ (dashed) at $T=0.5$; in the inset 
$\chi(C)$ for the same $t_w$'s, compared to FDT (dotted). Right: zoom
over the initial decay of $C$.}
\label{self-correlator}
\end{figure}

Richer information about the nonequilibrium evolution
is contained in the self-correlation, $C$, and linear self-response
$R$~\cite{Bou-review,Cuku}. 
An expression for $C$ in terms of 
$\Gamma$ that is also valid for the dilute model 
is given by Eq.~(2.12) in \cite{Cude}.
We identify several scalings. 
When $t$ and $t_w$ are long but shorter than 
$t_{\sc co}$ the dynamics resembles the one 
for the {\sc fc} model. For $T>T_0$ the decay is stationary.
For $T<T_0$ instead there is a first stationary 
decay towards an Edwards-Anderson 
plateau $q_{EA}=1-T/T_0$ when $\tau\equiv t-t_w\ll  t_w$ 
and a subsequent 
aging decay from $q_{EA}$ towards zero with 
simple aging scaling, $C(t,t_w) \sim {\cal C}(t_w/t)$,
when $t_w \propto t$. 
Instead, when $t_w$ falls in the tail-dominated regime 
$\Gamma(t)\sim f(t)$ and
\begin{eqnarray}
C(t,t_w) &\to&
\exp\left(-\frac{\tau^2}{8 p \ln t_w}\right)
\; 
\left[ 
1- 2T \int_0^{\frac{\tau}{2}} dy \, 
\exp\left(-\frac{y t_w}{p \ln t_w}\right) f(y)
\right]
\end{eqnarray}
when $t_w\to\infty$ and $\tau \ll  \ln t_w$. 
At $T=0$ we immediately identify the 
effective time $\tau_{\sc eff} \sim \sqrt{\ln t_w}$ 
and a very slow subaging decay~\cite{Be}. One can recast the correlation 
in the more familiar form
\begin{equation}
C(t,t_w) = \exp\left(-\frac{\ln^2 x}{8}\right) \; , \;\;
x=\frac{h(t)}{h(t_w)} \;, \;\; 
h(t) = \exp\left(\frac{t}{\sqrt{p \ln t}}\right)
\; .
\label{eq-scaling}
\end{equation}
If $\tau \ll  \sqrt{\ln t_w/t_w}$ we get $C\sim 1-T \tau$.
The duration of the $t_w$ independent part of the decay 
decreases with $t_w$ and eventually disappears. Thus,
the finite-$T$ corrections vanish for long $t_w$
and the scaling (\ref{eq-scaling}) holds for all $T$.
These features are shown in Fig.~\ref{self-correlator}.

The linear self-response is given by $R(t,t_w) =
\sqrt{\Gamma(t_w)/\Gamma(t)} f(\tau/2)$. 
When $t < t_{\sc co}$
the {\sc fc}  scaling holds. When $t_w \gg t_{\sc co}$
and for $\tau\propto \sqrt{\ln t_w}$  when $C$ varies,  
$R(t,t_w) = \exp[-\tau t_w/(2 p \ln t_w)]$.
$R$ decays much faster than $C$, it 
approaches zero on the  scale $\tau \sim \ln t_w/t_w$.
An equivalent
expression for this result is $R(t,t_w) \sim 
\exp [ (-\ln x) t_w/\sqrt{p \ln t_w} ]$
with $x$ given in Eq.~(\ref{eq-scaling}).

The modification of the fluctuation-dissipation theorem ({\sc fdt}) 
is different in the two 
non-equilibrium regimes. In the {\sc fc}-like scale the results
are identical to those in \cite{Cude} and the effective 
temperature~\cite{Cukupe} $T_{\sc eff} \equiv \partial_{t_w} C/R$ 
diverges as $\sqrt{t_w}$. In the tail-dominated scale $T_{\sc eff}$
diverges much faster, as 
$T_{\sc eff}[C] \sim \exp[a(C) t_w/\sqrt{p \ln t_w}]$
with $a(C) = \sqrt{(-2 \ln C)}$. The inset in 
Fig.~\ref{self-correlator} shows the parametric 
plot of the integrated response $\chi$ against $C$~\cite{Bou-review}
when $t_w$ is in the latter regime.

We solved (\ref{langevin}) in a rather simple way 
thanks to the quadratic nature of the 
model. Unfortunately, this calculation cannot be adapted to treat 
cases with higher order interactions as are models with soft spins.
With the purpose of developing a more generic approach
to the dynamics of dilute models 
we introduce a method to derive macroscopic equations for $C$ and $R$ 
from the dynamic generating functional $Z_d$. We only sketch the
method here, the details shall be given in a forthcoming
publication~\cite{Secu2}.
The idea is to take
benefit of the formal analogy between the static replica free energy
and the dynamical action expressed with a supersymmetric (SUSY) formalism.
Using the SUSY formulation of stochastic processes~\cite{susy} the
two-point functions are encoded in a correlator 
$N Q(a,b) = \sum_{i=1}^N 
[ \langle \Phi_i(a) \Phi_i(b) \rangle  ]_J$ between 
two superfields $\Phi_i(a)$. $a$ and $b$ include
time and two Grassmann coordinates $\theta,\overline\theta$.
We define the fraction of sites with super-field
$\Phi_i$ identical to a chosen value $\Phi$ for all coordinates, in
analogy with the replica order parameter for dilute systems~\cite{cphi}, 
\begin{equation}
c[\Phi]\equiv N^{-1} \sum_{i=1}^N \prod_a \delta(\Phi(a) -\Phi_i(a))
\; .
\end{equation}
After introducing it in $Z_d$ and averaging over 
disorder~\cite{Secu2} we obtain 
the saddle-point equation
\begin{equation}
c_{\sc sp}(\Phi) = 
\lambda \: \exp\left[-\frac{1}{2}  \int da \Phi(a) \left(- D_a^{(2)} + 
z(a) \right) \Phi(a) - \left. 
\frac{\delta H_{\sc eff}}{\delta c(\Phi)}
\right|_{c_{\sc sp}}
\right]
\label{c-eq}
\end{equation}
with $\lambda$ a normalization constant, 
$D_a^{(2)}$ a differential operator, $\int da$ the integration over
time and Grassmann coordinates, and
\begin{eqnarray*}
\exp(-N H_{\sc eff}) &=& 
\left[ \; \exp\left(-\int da  \; H_J[\Phi_i(a)] \right)\; \right]_J
\; .
\end{eqnarray*}
This equation determines $c[\Phi]$ for {\it any} 
model for which the saddle-point evaluation 
is exact. (For models without spherical constraint, one removes the
Lagrange multiplier $z(a)$). It plays the role of
Eq.~(20) in \cite{Secu} for the
replica order parameter $c_{\sc sp}(\vec \phi)$  introduced to compute 
$\rho(\mu  )$. It cannot be solved without assumptions, yet an
iterative resolution similar to that of~\cite{Bimo,Secu} taking into
account more and more precisely the fluctuations in the local geometry
of the interactions can be implemented. For the present model one
recovers the same results obtained in this
Letter~\cite{Secu2}. We expect that this approach will also be successful for
non-quadratic models for which an exact resolution is not possible. 

In short, we showed that the
dilute spherical spin-glass realizes 
explicitly part of the scenario expected to hold beyond the 
``mean-field picture'' of glassy dynamics. The main ingredient of the
model is the unbounded distribution of local connectivities,
regardless of the sign of the interactions.
In the glassy context this model has, however, some flaws. 
Its equilibrium energy density is not bounded and, 
strictly speaking, it only has a low-$T$ phase.
The dynamics in late epochs does not depend on $T$ and when 
$T$ is close but above $T_0$ 
the stationary correlations do not decay in two steps. 
Nevertheless, the existence of a dynamic crossover around $T_0$ as opposed
to a phase transition and the presence of two nonequilibrium 
regimes are expected features for real glasses. 

Dilute magnets are important in several other contexts. 
They are prototypical models  with Griffiths phases~\cite{Bray} and
heterogenous dynamics~\cite{Andrea}. They are also related to several
optimization problems, and their dynamics to that of optimization
algorithms~\cite{remi-review}. 
We shall report on studies of the dynamics of other dilute models,
hopefully free of the defects mentioned above, in the future.

\acknowledgments
The authors acknowledge financial support from an ACI grant and thank 
J. Kurchan for discussions. L.F.C. is associate researcher at ICTP Trieste
and a Fellow of the Guggenheim Foundation.

\end{document}